\documentstyle[psfig]{europhys}

\def\etal{{\hbox{{\tenit\ et al.\/}\tenrm :\ }}}

\def\And{{\rm and\ }}

\def\stars{\bigskip\centerline{***}\medskip}

\newif\ifboo \boofalse

\def\Review#1{\boofalse{\it #1},}
\def\Name#1{{\sc #1},}
\def\Vol#1{\ifboo Vol. {\bf #1}\else{\bf #1}\fi}
\def\Year#1{\ifboo #1\else(#1)\fi}

\def\Page#1{\ifboo {\rm p. #1}\else{\rm #1}\fi}

\begin{document}

\euro{xx}{x}{xx-xx}{xxxx}
\Date{}
\shorttitle{M. CAPONE \etal THE SMALL POLARON CROSSOVER 
ETC.}

\title{{\Large The small polaron crossover: 
comparison between exact results and
vertex correction approximation}}
\author{M. Capone\inst{1}\footnote{Present address: 
International School 
for Advanced Studies, Via Beirut 4, 34013 Trieste, Italy.},
S. Ciuchi\inst{2} and C. Grimaldi\inst{3}}
\institute{
 \inst{1} Dipartimento di Fisica, Universit\'{a} di Roma 
``La Sapienza".
P.le A. Moro 2, 00185 Roma, Italy \\
\inst{2} Dipartimento di Fisica, Universit\'{a} de L'Aquila,
via Vetoio, 67100 Coppito-L'Aquila, Italy and I.N.F.M., 
Unit\'a
de L'Aquila.\\
\inst{3} I.N.F.M., Unit\'{a} di Roma 1,
Dipartimento di Fisica, Universit\'{a} di Roma ``La 
Sapienza".
P.le A. Moro 2, 00185 Roma, Italy}

\rec{}{}

\pacs{
\Pacs{71}{38$+i$}{Polarons and electron-phonon interactions}
\Pacs{63}{20Kr}{Phonon-electron and phonon-phonon 
interactions}
      }

\maketitle

\begin{abstract}
We study the crossover from quasi free electron to small polaron
in the Holstein model for a single electron
by means of both exact and self-consistent calculations in
one dimension and on an infinite coordination lattice.
We show that the crossover occurs when both strong coupling
($\lambda>1$)
and multiphonon ($\alpha^2 >1$) conditions are fulfilled leading to
different relevant coupling constants ($\lambda$) in adiabatic
and ($\alpha^2$) anti adiabatic region of the parameters space.
We also show that the self-consistent calculations obtained
by including the first electron-phonon vertex correction give accurate
results in a sizeable region of the phase diagram well separated
from the polaronic crossover.
\end{abstract}

Recent optical measurements of the insulating parent 
compounds of the high-temperature superconductors \cite{htc} 
show the presence of polaronic carriers, and 
evidence for strong electron-phonon (el-ph) coupling 
effects has been given also for the 
colossal magnetoresistance manganites \cite{mang} and Nickel 
compounds \cite{calvani}.
These findings underline the necessity of a clear 
theoretical description of electron-phonon coupled system 
and more specifically of the constraints for the existence of 
the small polaron ground state.
This state, characterized by strong local electron-lattice 
correlation, is definitively a non-perturbative phenomenon,
and cannot be described by simple summation of the 
perturbative series such as the one which defines the 
Migdal-Eliashberg (ME) theory \cite{me,marsiglio}.

The aim of this work is to provide a detailed study of the 
crossover which occurs at intermediate electron-lattice 
couplings from quasi-free electron to small polaron ground state. 
We also study the role of the lattice dimensionality and compare 
exact results with self-consistent theories.

A single electron interacting with Einstein phonons through
an Holstein type local interaction is the simplest system 
which shows such kind of crossover. The associated 
hamiltonian is 
\cite{holstein}:

\begin{equation}
\label{holham}
{\cal H}=-t\sum_{\langle ij\rangle}
c^\dagger_i c_j +
g\sum_i
n_i\left( a_i+a^\dagger_i \right)
+\omega_0 \sum_i a^\dagger_i a_i
\end{equation}
where $c_i$ ($c_i^{\dagger}$) is the destruction 
(creation) operator for an electron  on site $i$ , 
and $n_i=c^\dagger_i c_i$.
$a_i$ ($a_i^{\dagger}$) is the destruction
(creation) operator for Einstein dispersionless phonons with
frequency 
$\omega_0$ on site $i$.
The hamiltonian (\ref{holham}) represents
a non-trivial many-body problem even in the single electron 
case
due to the quantum nature of phonons
and it has been already studied in recent years 
by means of numerical 
\cite{marsiglio,deraedt,demello,fehske} 
and analytical \cite{alexandrov,lavorone,zhao} techniques.

For the hamiltonian of eq.(\ref{holham}) two dimensionless 
parameters, which measure the
electron-lattice coupling, are introduced: 
$\lambda=g^2/(D\omega_0)$ and $\alpha=g/\omega_0$,
where $D = 2td$ is the 
half-bandwidth
for the free electron
and $d$ is the system dimensionality.

$\lambda$ is originally introduced in the standard weak 
coupling pertubation theory ($g/t\ll 1$) and 
is the coupling parameter of a 
ME approach in the case of one electron.
On the other hand $\lambda$ is the ratio 
between the small polaron energy
$E_{\mbox{p}} = -g^2/\omega_0$ and the free electron energy 
$E_{\mbox{free}} = -D$, so 
it naturally measures the energetic gain of the small 
polaron state with respect to the free electron-like state.

The parameter $\alpha$ is instead introduced in the standard small polaron 
theory and is also the relevant coupling in the  {\it atomic
limit} ($t=0$). In this limit $\alpha$ measures the lattice 
displacement associated to the polaron and 
$\alpha^2$ is the average number of phonons bound to the 
electron.
According to
the Lang-Firsov results \cite{lf}
followed by the Holstein approximation\cite{holstein} it 
also rules the reduction of the effective
hopping $t^{*} = t\exp{(-\alpha^2)}$ \cite{capone,ciuchi}.

Besides $\lambda$ and $\alpha$, the el-ph system described by
eq.(\ref{holham}) is governed also by another dimensionless 
parameter: $\omega_0/t$. It measures the 
the degree of adiabaticy of the lattice motion (lattice kinetic 
energy $\simeq \omega_0$)
compared to the electron one (electron kinetic energy 
$\simeq t$)\footnote{We stress that all the parameters we 
consider are
defined in terms of
the {\it bare} quantities $t$, $\omega_0$ and $g$ appearing
in the hamiltonian (\ref{holham}).}.

A bound state between electron and phonon is formed as soon 
as $\lambda > 1$. In the adiabatic regime ($\omega_0/t \ll 
1$) this condition is sufficient to give a polaronic state
since the electron is bound to the slowly moving lattice giving
rise to a strong enhancement of effective mass.
In the antiadiabatic regime ($\omega_0/t \gg 1$) such a
picture is no longer true due to the fast lattice motion.
In this case, polaronic features such as strong local 
electron-lattice correlations arise only when the electron is
bound to a {\it large} number of phonons.
This condition is fulfilled for $\alpha^2 > 1$. 
To summarize in both adiabatic and
antiadiabatic regimes to have a polaronic state we must have
{\it both} $\lambda > 1$ and $\alpha^2 > 1$ \cite{capone}.
The above discussion stresses that $\lambda > 1$ is not the
only condition for small polaron formation, 
in contrast with the claim of ref.\cite{alexandrov}.

The parameter $\omega_0/t$ influences also the dependence
of the behavior of the el-ph coupled system on the system
dimensionality.
We shall show that in the antiadiabatic regime the 
constraint for the small polaron state is rather
universal, 
{\it i. e.}, it does not depend on the system dimensionality.
On the other hand, dimensionality plays an important role in 
the adiabatic
limit $\omega_0/t = 0$. In fact, In d=1 the ground state is localized 
for any finite value of $\lambda$ and 
a crossover occurs between large and 
small polaron around $\lambda\simeq 1$ whereas for $d \ge 2$ it has
been shown that a localization transition occurs at finite $\lambda$
from free electron to small polaron\cite{kabanov}.

The relevance of the adiabatic parameter
$\omega_0/t$ and the role of dimensionality
is exploited non perturbatively
by using two alternative procedures, which both
give exact
numerical results:

i) Exact diagonalization of small one dimensional clusters 
by means
of the Lanczos algorithm (ED-1d).

ii) Dynamical mean field theory (DMFT-3d).

In the exact diagonalization approach, 
the infinite phonon Hilbert space has to be truncated to 
allow for a given maximum number of phonons per site 
$n_{\mbox{max}}$.
In order to properly describe the multiphonon regime (expecially
in the adiabatic regime where a large number of low energy 
phonons can be excited) our cut-off is $n_{\mbox{max}} =20$. 
This relatively
high value forced us to restrict to a four-site cluster with
periodic boundaries condition in the strong-coupling 
adiabatic regime. In the weak-coupling regime and for 
larger phonon frequencies
a lower value of $n_{\mbox{max}}$ is needed, allowing us to 
consider larger clusters up to ten sites.
We checked that finite-size effects do not significantly 
affect the cross-over coupling, since small-polaron formation is a
local, high energy process.

The dynamical mean field theory approach
can be seen as the exact solution of the small polaron  problem
defined on an infinite coordination lattice. For this reason 
this theory does not suffer of limitations of other approach such 
as the variational \cite{ciuchi} which may be in contraddiction 
with the Gerlach-Lowen theorem  \cite{lowen,ciuchi2}.
The formulation of the DMFT requires the knowledge of the 
free particle DOS so that  by choosing a semi-circular
free particle DOS, it is possible to mimic a realistic 
three-dimensional case (DMFT-3d). 
Details of perturbation theory expansion
in the DMFT framework are given in Ref. \cite{lavorone}
together with results concerning the exact spectral properties.

Here, we study the behavior of the ground state energy $E_0$
using the exact
solutions ED-1d and DMFT-3d  and we compare the results with 
the 
self-consistent non-crossing (NCA) 
and vertex corrected approximations (VCA).
These two approximations are defined by the self-consistent 
calculation of the electronic zero-temperature 
self-energy $\Sigma(k,\omega)$ given below:

\begin{equation}
\label{selfen}
\Sigma(k,\omega) \!= \! \frac{2\lambda\omega_0 
t}{N}\sum_{p}G(p,\omega-\omega_0) 
\left[1\! +\!\frac{2\lambda\omega_0 t}{N}\sum_{q}G(q-
p+k,\omega-\omega_0)
G(q,\omega-2\omega_0)\right] ,
\end{equation}
where $G(k,\omega)$ is the retarded fully renormalized  
single electron 
Green's function:

\begin{equation}
\label{green}
G(k,\omega)^{-1}=\omega-\epsilon_k-\Sigma(k,\omega)+
i\delta .
\end{equation}
which will be determined self-consistently.
The NCA approach amounts to compute $\Sigma$ by
retaining only the $1$ in the square brackets of eq. 
(\ref{selfen}). 
NCA is formally similar to the ME approximation
for metals but it has to be stressed 
that Migdal criterion has no sense in the case of only one 
electron 
having a vanishingly small Fermi surface. 
The VCA is given by the inclusion also of the second term 
in square brackets of eq.(\ref{selfen}) which represents
the first vertex correction.
This approach is formally similar to the approximation scheme 
used in the formulation
of the non-adiabatic theory of superconductivity 
\cite{grimaldi}. The
present calculations provide therefore also a test of reliability 
of such an approximation for the one-electron case.
The evaluation of self-energy allows to compute the ground
state energy
given by the lowest energy pole of eq.(\ref{green}).
In the context of dynamical mean-field theory the internal 
propagators 
appearing  in eq. (\ref{selfen}) are averaged over the 
$k$-space\cite{ciuchi} and the self-energy turns out to be 
$k$-independent
at any perturbative order. 

\begin{figure}
\vbox to 8.3cm{\vfill
\centerline{\psfig{figure=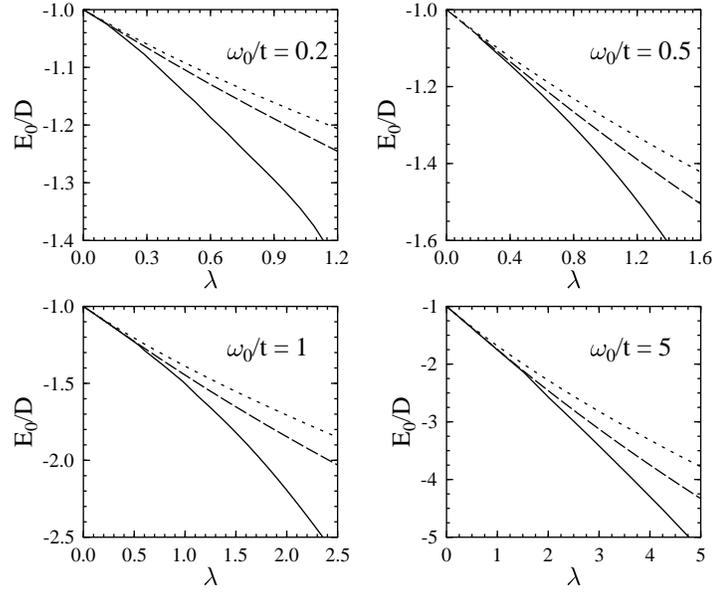,width=8cm}}\vfill}
\caption{Ground state energy results in d=1. The exact 
diagonalization 
results are compared with the NCA (short dashed) and VCA 
(long dashed)
calculations.}
\label{fig-energia-d1}
\end{figure}
\begin{figure}
\vbox to 8.3cm{\vfill
\centerline{\psfig{figure=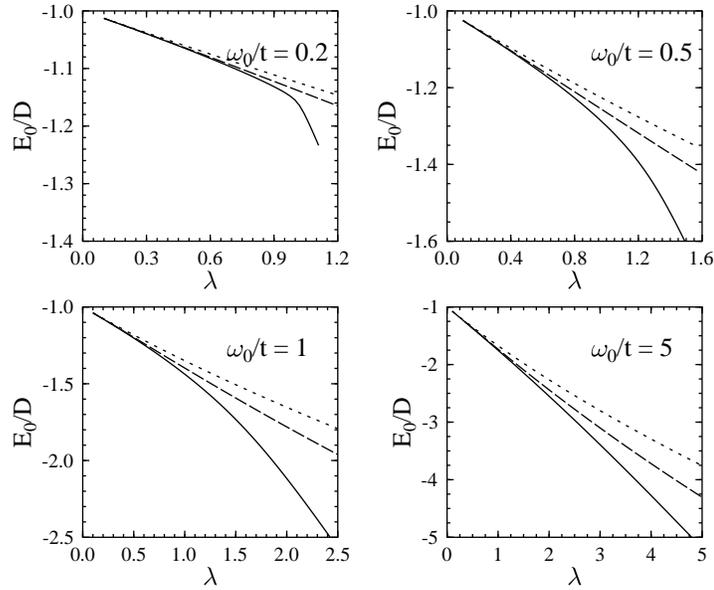,width=8cm}}\vfill}
\caption{Ground state energy results for an infinite 
coordination
lattice. Comparison between
dynamical mean field (solid line), NCA (short dashed) and 
VCA (long dashed).} 
\label{fig-energia-d3}
\end{figure}

In fig. \ref{fig-energia-d1} we compare the ground-state energy $E_0$ obtained by ED-1d\footnote{Different cluster sizes and values of 
$n_{max}$ have been used in the different physical regimes 
in order to minimize the finite-size and phonon cut-off effects.} 
with the NCA and VCA results. 
The same quantities evaluated in the DMFT-3d case are shown 
in fig. \ref{fig-energia-d3}.
We have chosen the same half-bandwidth $D$ in both DMFT-3d 
and ED-1d cases.

In the adiabatic regime the agreement of both approximations
with exact results strongly depends on the system 
dimensionality
as a result of the different low-energy behaviour of the 
DOS. 
In fact, moving from $\omega_0/t=0.2$ to $\omega_0/t=0.5$ 
the agreement of 
the self-consistent calculations with the exact results is 
improved for
the 1d case (fig. \ref{fig-energia-d1}) whereas it becomes poorer for the 3d 
one (fig. \ref{fig-energia-d3}).
Both approximate and exact results
tend to become independent on the dimensionality as far as
$\omega_0/t$ is increased as it is seen from the comparison
of fig. \ref{fig-energia-d1} and \ref{fig-energia-d3} 
for large $\omega_0/t$.
This can be undestood in terms of scattering process
which in the anti-adiabatic case will
lead electrons through intermediate states
out of the band. In this scattering process the system can
be thought as a flat band ``atomic" system in interaction with 
high energy phonons.
However, the VCA approach represents a significative 
improvement with respect
to the non-crossing approximation for every system 
dimensionality and over
a wide range of parameters. It is also clear from figs. \ref{fig-energia-d1} 
and \ref{fig-energia-d3} that
both the self-consistent NCA and VCA calculations deviate
from the exact results when the crossover towards the small 
polaron regime is approached. 

\begin{figure}
\vbox to 6cm{\vfill
\centerline{\psfig{figure=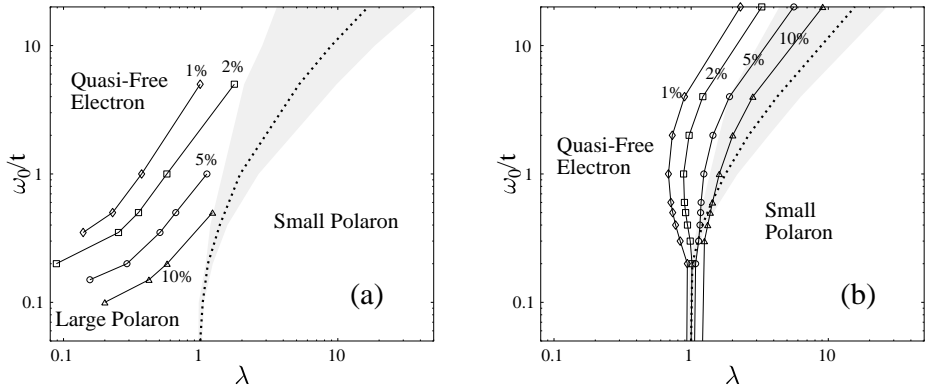,width=13cm}}\vfill}
\caption{Phase diagram in the $\lambda$-$\omega_0/t$ plane 
for the one-dimensional (a)
and the infinite coordination lattice (b) Holstein model.
The dotted line is the polaron crossover value $\lambda_c$ 
and the width of the crossover is 
evidentiated by a shaded area. Notice that the crossover is much
broader in the antiadiabatic regime compared to the adiabatic one.
The isolines represents the relative difference 
between the exact and the VCA result for the ground state 
energy.}
\label{fig-fase-d1}
\end{figure}

An exhaustive study of the comparison between 
the exact results and the VCA approach in the parameter
space $\lambda$-$\omega_0/t$ is shown in figs. \ref{fig-fase-d1}(a)-(b). 
We explicitly evaluated both in 1d and 
3d the relative difference 
$\delta E_0  = 2|E_0^{\mbox{VCA}} - E_0^{\mbox{exact}}|/
|E_0^{\mbox{VCA}}+E_0^{\mbox{exact}}|$ 
where $E_0^{\mbox{exact}}$ and  $E_0^{\mbox{VCA}}$ are the 
ground-state energies evaluated
by exact techniques and the vertex-corrected approximation, 
respectively.
To analyze the region in the parameter space where the VCA agrees within 
a given accuracy with the exact results 
we report lines of constant $\delta E_0$. 

The agreement between self-consistent approximations and 
exact results is sensible to system dimensionality.
In dimensions larger than two approaching the adiabatic 
limit and for
small couplings the electron tends to be free.
For this reason self-consistent approximations work well.
On the contrary in the adiabatic limit and for d=1 the ground state 
is a large polaron and self-consistent approximations fail to 
predict its energy.
In general,
self-consistent approximations work well outside the
polaron region whatever polarons are either small or large.
This can be seen directly from figs. \ref{fig-fase-d1}(a)-(b) 
where the critical 
coupling $\lambda_{\mbox{c}}$ of the crossover 
to small polaron  is depicted as a dotted line. 
The critical coupling $\lambda_c$  is defined as the value at which
$d E_0/d g$ has maximum slope.
By Hellmann-Feynman 
theorem $d E_0/dg$ is just the electron lattice local
correlation function $\langle n_i
(a_i+a^{\dagger}_i)\rangle$. In the same figures, we provide
also an estimate of the width of the crossover (shaded areas)
obtained by looking at
the maximum slope of $|\partial^2 E_0/\partial g^2|$.
We checked that different criteria, like {\it e. g.} 
the effective mass enhancement\cite{ciuchi}, 
provide the same qualitative results.

In conclusion, we have shown that the crossover toward
the small polaron state depends strongly on the adiabaticity
parameter $\omega_0/t$. In the antiadiabatic regime
the crossover is ruled by $\alpha^2$ and it is independent
of the system dimensionality whereas in the adiabatic regime
the relevant coupling is $\lambda$ and the details of the
crossover depend on the dimensionality.
We have also shown that self-consistent calculations
provide ground state energies which agree well with
exact results in the quasi free electron regime and
that such an agreement
is increased when vertex corrections are
taken into account.   

\stars{We thank M. Grilli, F. de Pasquale, D. Feinberg and 
L. Pietronero for stimulating
discussions. C. G. acknowledges the support of a I.N.F.M. 
PRA project.}

\end{document}